\documentclass[nature, twocolumn, superscriptaddress]{revtex4-1}
\usepackage{epsfig}
\usepackage{amssymb}
\usepackage{color}
\begin{document} 

\title{Anomalous dielectric response at intermixed oxide heterointerfaces} 

\author{Valentino R. Cooper}\email{coopervr@ornl.gov}\thanks{\\This manuscript has been written by UT-Battelle, LLC under Contract No. DE-AC05-00OR22725 with the U.S. Department of Energy. The United States Government retains and the publisher, by accepting the article for publication, acknowledges that the United States Government retains a non-exclusive, paid-up, irrevocable, world-wide license to publish or reproduce the published form of this manuscript, or allow others to do so, for United States Government purposes. The Department of Energy will provide public access to these results of federally sponsored research in accordance with the DOE Public Access Plan.\\} \affiliation{Materials Science and Technology Division, Oak Ridge National Laboratory, Oak Ridge, TN 37831}
\author{Houlong L. Zhuang} 
\affiliation{Center for Nanophase Materials Sciences, Oak Ridge National Laboratory, Oak Ridge, Tennessee 37831, USA} 
\author{Lipeng Zhang} 
\affiliation{Department of Materials Science and Engineering, The University of Tennessee, Knoxville, Tennessee 37996, USA} 
\author{P. Ganesh} 
\affiliation{Center for Nanophase Materials Sciences, Oak Ridge National Laboratory, Oak Ridge, Tennessee 37831, USA} 
\author{Haixuan Xu} 
\affiliation{Department of Materials Science and Engineering, The University of Tennessee, Knoxville, Tennessee 37996, USA} 
\author{P. R. C. Kent} 
\affiliation{Center for Nanophase Materials Sciences, Oak Ridge National Laboratory, Oak Ridge, Tennessee 37831, USA} 
\affiliation{Computer Science and Mathematics Division, Oak Ridge National Laboratory, Oak Ridge, Tennessee 37831, USA} 
\date{\today}

\maketitle
\textbf{Two-dimensional charge carrier accumulation at oxide heterointerfaces presents a paradigm shift for oxide electronics. Like a capacitor, interfacial charge buildup couples to an electric field across the dielectric medium. To prevent the so-called polar catastrophe, several charge screening mechanisms emerge, including polar distortions and interfacial intermixing which reduce the sharpness of the interface.  Here, we examine how atomic intermixing at oxide interfaces affect the balance between polar distortions and electric potential across the dielectric medium. We find that intermixing moves the peak charge distribution away from the oxide/oxide interface; thereby changing the direction of polar distortions away from this boundary with minimal effect on the electric field. This opposing electric field and polar distortions is equivalent to the transient phase transition tipping point observed in double well ferroelectrics; resulting in an anomalous dielectric response -- a possible signature of local negative differential capacitance, with implications for designing dissipationless oxide electronics.}
 
Advances in thin film growth techniques, which allow for precise layer-by-layer growth of epitaxial materials, have opened the door to numerous discoveries, which have not been achievable through bulk synthesis techniques. A prominent example is the emergent phenomena\cite{Okamoto04p630} reported at $AB$O$_3$ perovskite oxide heterointerfaces such as 2DEGs,\cite{Ohtomo02p378, Ohtomo04p423, Shibuya04pL1178, Hwang12p103, Zubko11p141, Mannhart10p1607} interfacial superconductivity\cite{Reyren07p1196} and novel magnetic properties.\cite{Brinkman07p493, Glavic16p140413} However, despite the unprecedented control, it is known that some interfaces are inherently prone to interfacial intermixing and thus the production of clean/sharp interfaces may not be possible. Nakagawa and co-workers demonstrated that in LAO/STO oxide heterostructures the potential created across a SrO-TiO$_2$-LaO interface due to the difference in La and Sr charges can be partially compensated for by interfacial intermixing.\cite{Nakagawa06p204} Such effects may significantly define electronic properties at these interfaces. For example, theory has demonstrated that interfacial intermixing could alter the displacement of ions or polarization at an interface.\cite{Cooper07p020103R, Chen16p104111} This change could have consequences for the way that charge is screened at the interface. Similarly, work on $\delta$-doped oxide superlattices have shown that a decrease in the fraction of La in the interfacial layer can have positive effects on the mobility of electrons at the interface.\cite{Choi12p4590} As such,  the ``sharpness" of an interface has important implications for the modulation of charge carrier densities and mobilities.

\begin{figure}[ht]
\includegraphics[width=2.25in]{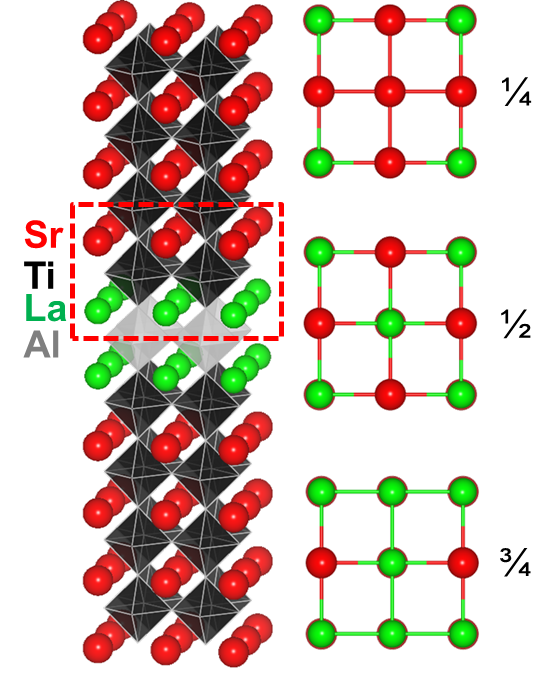}
\caption{\label{fig1} Structural model of the superlattice used to explore the effects of interfacial intermixing on the carrier densities and mobilities at the LAO/STO interface. The model emphasizes the $\delta$-doped interface. Red and green spheres represent Sr and La ions, respectively. Black and grey octahedra indicate TiO$_6$ and AlO$_6$ units, respectively. 2D structures to the right indicate the intermixing in the SrO layer for 1/4, 1/2 and 3/4 intermixing. In each case, the adjacent LaO layer is the inverse of this layer. }
\end{figure}

In this article, we examine the effect of $A$-site interfacial intermixing at 1.5 LAO/8.5 STO oxide heterostructure interfaces (as illustrated in Fig.~\ref{fig1}) on carrier densities and mobilities in La $\delta$-doped SrTiO$_3$ using first-principles density functional theory. Our calculations show that intermixing above 1/4 concentration is unstable relative to clean interfaces (once configurational entropy is considered); thus having implications for large scale production, where experimental growth techniques, such as chemical vapor deposition, may be a viable alternative.  However, it is worth noting that larger intermixing fractions have been stabilized using epitaxial deposition techniques like pulsed laser epitaxy (PLE).\cite{Choi12p4590} Furthermore, we predict noticeable reductions in interfacial 2DEG carrier densities with increasing intermixing fraction, albeit the total excess charge density remains 0.5 e$^-$/interface. We further demonstrate that intermixing is correlated with decreases in the band effective masses of the $t_{\rm 2g}$ states that give rise to the 2DEG at the interface. As suggested by previous studies on fractionally $\delta$-doped interfaces\cite{Choi12p4590} and potassium-based 2DEG systems,\cite{Cooper12p235109, Shen15p74, Zou15p36104} this reduction in band effective masses signals possible enhancements in electron mobilities. These results offer a plausible explanation for the deviations in carrier mobilities and densities measured in different experimental samples.\cite{Seo07p266801} Interestingly, together these effects result in unusual response in the static dielectric constant at the interface--resulting in the possibility of a negative dielectric constant for intermixing fractions of greater than 1/2. This phenomenon can be explained through a simple electrostatic model similar to that used to explain the observance of negative differential capacitance at a capacitor/ferroelectric interface\cite{Catalan15p137, Gao14p5814, Khan15p182, Zubko16p524}. Ultimately, this work may lead to novel routes to controlling emergent phenomena at oxide interfaces.   

\begin{table}
\caption{\label{tab1} Enthalpy of intermixing, $\Delta$ E$_{\rm mixing}$ [meV/interface u.c.], relative to the abrupt interface and the fractional $d_{xz}+d_{yz}$ occupancy of 2DEG states, $f$, for the  intermixed interfaces studied.}
\begin{tabular}{ c  c c }
\hline
Intermixing & $\Delta E_{\rm mixing}$ & $f_{d_{xz}+d_{yz}}$ \\
    \hline			
  abrupt & - - - & 0.245\\
   $\frac{1}{4}$ & 4 & 0.270\\
   $\frac{1}{2}$ & 59 & 0.258\\
   $\frac{3}{4}$ & 85 & 0.238\\
   \hline  
   
\end{tabular}
\end{table}

\begin{figure}[t]
\includegraphics[width=3.25in]{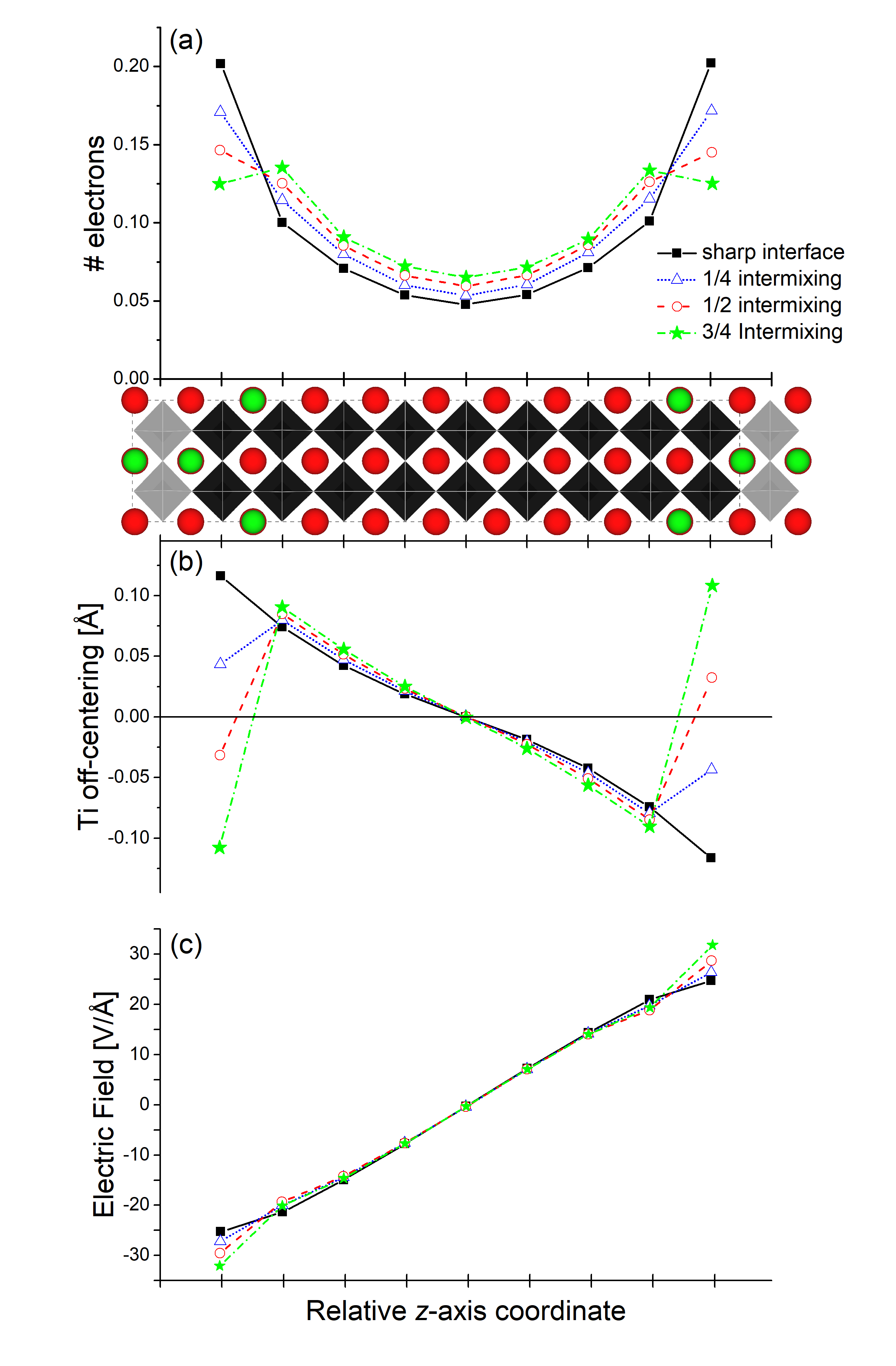}
\caption{\label{fig3} (top) Representative 2D projection of the relaxed atomic displacements in the abrupt interface LAO/STO superlattice. (a) Charge distribution, (b) $A$-site and (c) $B$-site cation off-center displacements (polar distortions) as a function of $z$ coordinate relative to the AlO$_2$ plane for the superlattices studied. Solid black (circles), red dotted (square), blue dashed (diamonds) and yellow dot-dashed (triangles) lines represent the abrupt, 1/4, 1/2 and 3/4 intermixed interfaces, respectively.}
\end{figure}

\textbf{\emph{Interface stability.}} To understand the tendency towards intermixing, we first computed the enthalpy of intermixing $\Delta E_{\rm mixing}$ for the 1/4, 1/2, and 3/4 intermixed interfaces in superlattices comprised of 8.5 layers of STO and 1.5 layer of LAO (i.e.~-LaO-AlO$_2$-LaO-TiO$_2$-SrO-TiO$_2$-SrO-TiO$_2$-SrO-TiO$_2$-SrO-TiO$_2$-SrO-TiO$_2$-SrO-TiO$_2$-SrO-TiO$_2$-SrO-TiO$_2$-). (N.B. intermixing was considered on both sides of the interface). Table~\ref{tab1} lists the intermixing energy relative to the abrupt/sharp interface as a function of intermixing fraction. For the 1/4 intermixing fraction, we find that the intermixing energy is only 4 meV/interface u.c. higher than the abrupt interface. This is well below the thermal energy at room temperature. Indeed, a careful consideration of the configurational entropy of mixing shows that at room temperature this intermixing fraction would be stabilized over that of the abrupt interface. However, as we increase the intermixing fraction to 1/2 and 3/4 we find that the interface is severely destabilized (more than could be recovered through configurational entropy alone). As such, our calculations imply that the maximum interfacial intermixing fraction will be on the order of 25\%. This is consistent with experimental electron energy loss spectroscopy (EELS) observations of the La distribution in the prototypical LAO/STO system which indicate a roughly 25-30\% of La in the SrO layer.\cite{Cantoni12p3952}

\begin{figure}[t]
\includegraphics[width=3.5in]{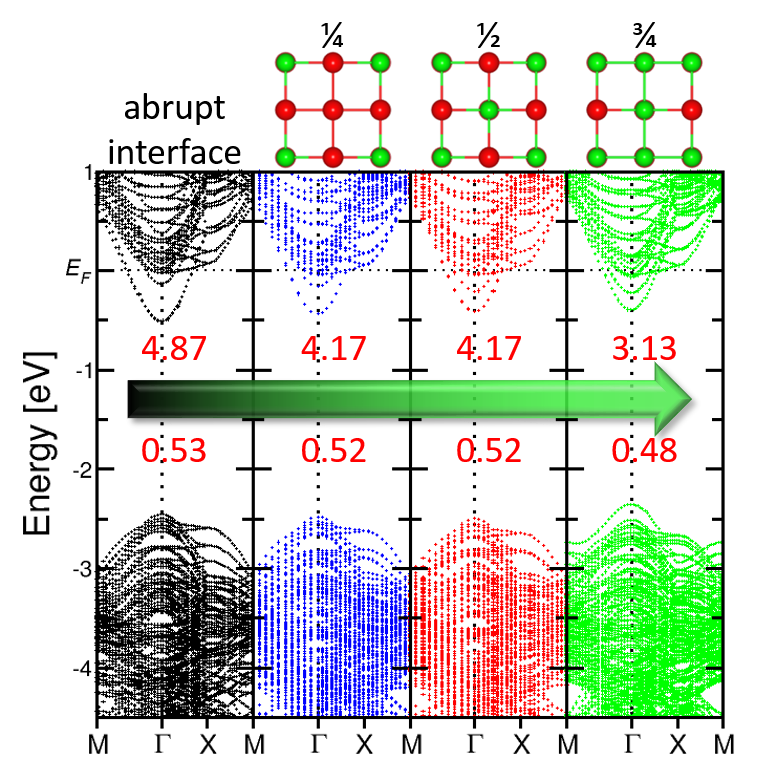}
\caption{\label{fig4} Electronic band structure for (a) abrupt, (b) 1/4, (c) 1/2 and (d) 3/4 intermixing emphasizing the partially occupied states near the Fermi level. Numbers above the arrow are the band effective masses of the occupied heavy mass band going in the $\Gamma$-X direction. Numbers below the arrow are the average band effective masses of the occupied parabolic bands around $\Gamma$. }
\end{figure}

\textbf{\emph{Electrostatics, structural distortions and charge carrier densities.}} Previous work suggests that the electrostatic potential across an interface, arising from the difference in charge between Sr$^{2+}$ and La$^{3+}$, may drive interfacial cation intermixing.\cite{Nakagawa06p204} This potential further induces compensating polar distortions away from the interface, which functions to  screen the accumulated charge at the interface. In a recent review, Bristowe {\it et al.} hypothesized that the effect of interfacial intermixing would be to merely shift the origin of the electrostatic potential.\cite{Bristowe14p143201} Such a change would have specific consequences on the displacement patterns (involved in screening the interfacial charge) as well as the charge density distribution at the interface.  

Figure~\ref{fig3}b depicts the layer-averaged off-center displacements of the $B$-site Ti cations (i.e. the polar atomic distortions). Here, we see that for the superlattices with sharp interfaces there is an almost linear distribution of cation displacements away from the interfaces -- tending to zero displacement in the center of the superlattice  in the bulk-like STO layers. Interestingly, in the interfacial intermixing superlattices we observe two key features. First, as the magnitude of the interfacial intermixing increases there is an almost linear decrease in the magnitude of the Ti polar distortions in the first interfacial layer, with systems comprised of $>$ 50\% intermixing having displacements in the opposite direction as the neighboring layers. Second, the magnitude of the displacements of Ti cations in the remaining layers retains nearly the same as the corresponding layers in the sharp interface system.   

\begin{figure}[b]
\includegraphics[width=3.45in]{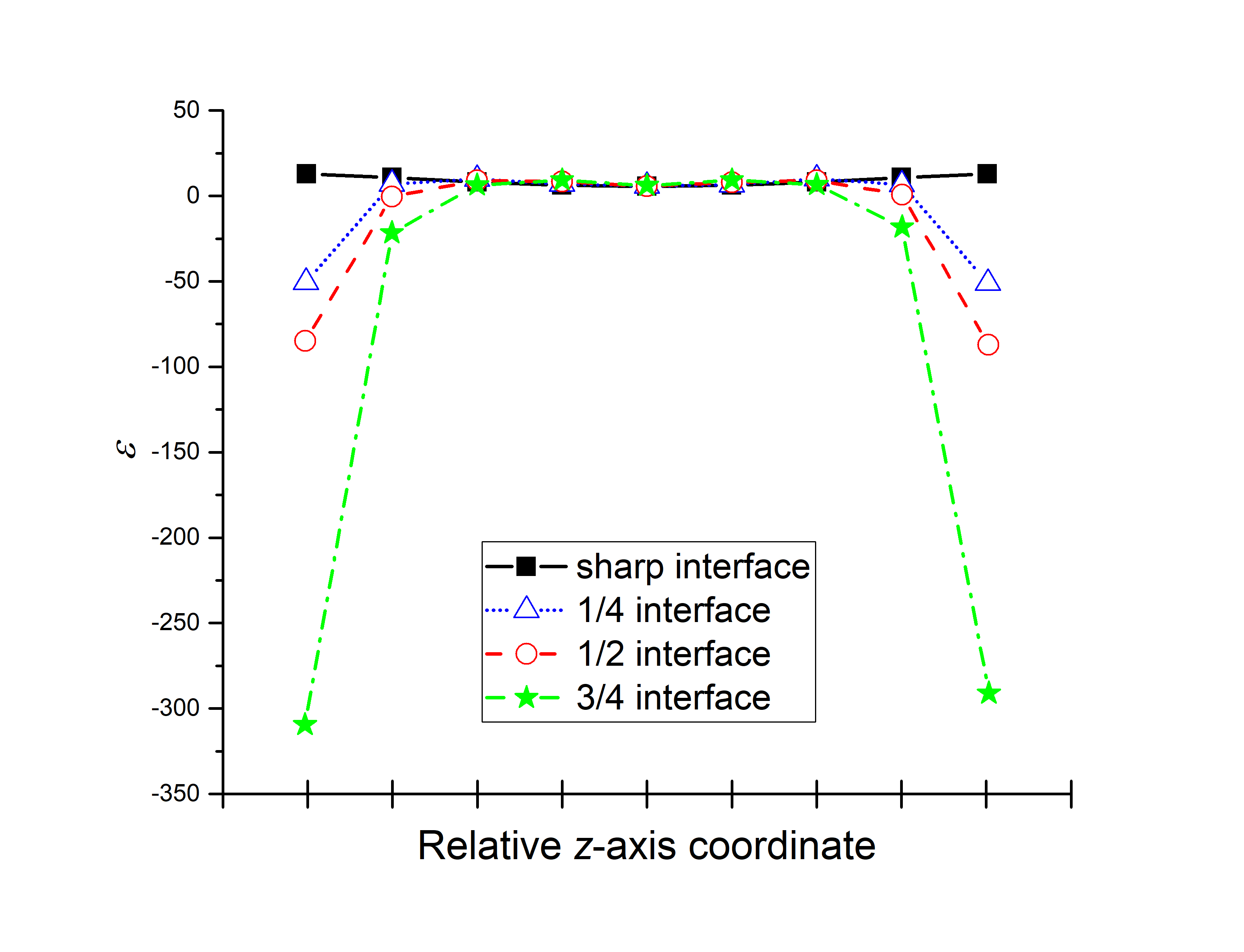}
\caption{\label{fig3b} Dielectric permittivity, $\epsilon$, computed based on the polar distortions and electric field due to the electrostatic potential as a function of $z$-axis coordinate relative to the AlO$_2$ plane for the superlattices studied. Solid black (squares), blue dotted (triangles), red dashed (circles), and green dot-dashed (stars) lines represent the abrupt, 1/4, 1/2 and 3/4 intermixed interfaces, respectively.}
\end{figure} 

Such displacements are typically correlated with changes in the distribution of charge in the interfacial layers. To compute the charge density due to the interfacial 2DEG states, we integrate the partial density of states (PDOS) from the Fermi energy to $\sim$-2~eV below (see band structure in Fig.~\ref{fig4}). (N.B. These are comprised of only Ti $t_{\rm 2g}$ states.) Figure~\ref{fig3}a depicts the layer-by-layer average Ti $t_{\rm 2g}$ states giving rise to the 2DEG charge density. We observe that as the intermixing fraction is increased there is a gradual broadening of the interfacial charge density distribution. Commensurate with this broadening, there is a reduction in peak charge density. For intermixing of less that 1/2, there is no shift in the peak charge density away from the initial interface layer; thereby implying a softening of the interfacial electrostatic potential in the vicinity of the intermixed layers. Conversely, for the 3/4 interface there is a 38\% reduction in charge at the interface and the peak charge density now shifts one layer further away from the initial interface. From there, the reduction in the electron count is again similar to the abrupt layer. 

To better understand this behaviour we examine the electric field across the STO region of the superlattice. Figure~\ref{fig3}c displays the electric field computed from the layer averaged electrostatic potential (minus the bare ion potential) for the four systems studied. Here, at the interface we see significant enhancements in the interfacial electric field with increases in the magnitude of the electric field as large as 32\% relative to the sharp interface in the case of 3/4 intermixing. However, we find that in the bulk of the superlattice intermixing has no effect on the magnitude of the electric field. In essence, 3/4 intermixing creates a new, almost abrupt La/Sr interface, which shifts the electrostatic potentials by one layer; thereby confirming the previous speculation by Bristowe et al.\cite{Bristowe14p143201}

\begin{figure*}[t]
\includegraphics[height=3.5in]{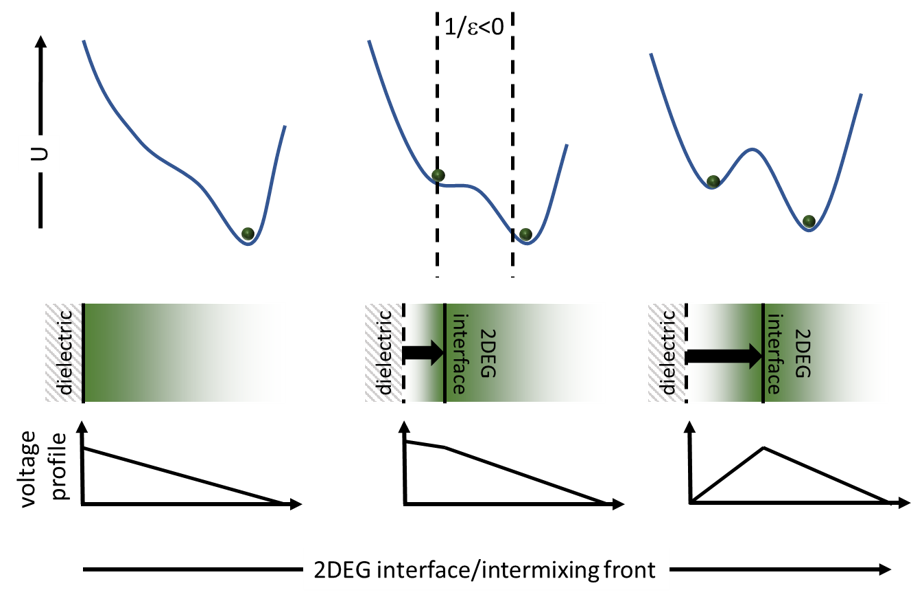}
\caption{\label{fig5} Schematic of the dielectric response of an intermixed system. (top) Evolution of the energy landscape as a function of movement of the intermixing front i.e. going from a clean interface to one in which the LaO layers are decoupled from the AlO$_2$ dilectric layer. (middle) Schematic of change in polar distortions away from 2DEG interface, dark green areas indicate regions of highest polarization. (bottom) Volatage profile across the STO medium. For intermixing close to the dielectric AlO$_2$ layer, there is little effect on the voltage drop across the bulk, as the 2DEG interface layer moves to the center of the bulk material, we can expect the voltage drop to recover so as to match the direction of the polarization. }
\end{figure*}

The crossover in the atomic displacement patterns and charge density distribution at around 1/2 intermixing demonstrates the dependence of the electrostatic/physical behaviour at the interface on the chemical makeup of the interface. In the case of partial mixing (i.e. $<$ 50\%),  this results in a softening of the charge layer imbalance at the interface, while for large mixing fractions the interface is essentially moved one layer over; again leading to the physically predicted picture of shifting the origin of the interfacial potential.\cite{Bristowe14p143201}

\textbf{\emph{Dielectric Response.}}
One unique consequence of the interfacial intermixing is the fact that the polar distortions change direction, opposing that of the local electric field. If one considers the ferroelectric superlattice picture where the displacement field, $D$, remains a constant, this would suggest changes in the dielectric constant, $\epsilon_0$ of the material. Figure~\ref{fig3b} depicts the computed local dielectric constant that would be computed by this approach (see Methodology section) in a ferroelectric superlattice for these materials. Dramatically, we see that for all intermixed systems this would predict a negative dielectric constant at the interface, which quickly returns to the sharp interface constant value for the middle of the superlattices. With the caveat  that the interfacial region is charged and thus the traditional concept of polarization would not hold in this material. 

Figure~\ref{fig5} depicts a schematic representation of the evolution of the potential energy well and voltage profile across the superlattice as a function of position of the 2DEG interface position (or intermixing). An important distinction to make is that the calculation setup can be recast as one in which the LaO layer, which ultimately controls the position of the 2DEG interface, is moving away from the dielectric interface i.e. the AlO$_2$ interfacial layer. This has two fundamental consequences. The first being that the position of the peak 2DEG charge density defines the direction of the polar distortions (always away from this interface). As such, in both the 1/2 and 1/4 intermixing case, this density moves to 1 layer away from the AlO$_2$ layer. as such it fosters polar distortions away from this layer towards the AlO$_2$ layer; opposite to the polar distortions on the other side of the 2DEG peak charge interface. This movement of polar distortions is similar to the movement of a polarization domain wall in a ferroelectric and has natural correlations to previous works looking at ferroelectric capacitors sandwiched between a dielectric medium.\cite{Gao14p5814} Here, the shift in peak density can be thought to stabilize the transition state between the two potential wells. The second consequence is that coupled with the polar distortions there will be a change in the electric field drop across the slab. However, in the case of La/SrO intermixing near the AlO$_2$ layer, this has a minimal effect and the electric field direction is still largely controlled by the potential drop across the entire slab. We anticipate that this effect is only for small displacements of the 2DEG interface away from the AlO$_2$ interface and for movements of the 2DEG interface a few layers away the voltage profile will then again follow that of the polar distortions. In any event, it is the difference in distortions away from the 2DEG interface as well as the minimal effect on the direction of the electric field which stabilizes the anomalous dielectric response near the interface.  Such an analysis may be suggestive of novel physics at the interface. For example, a negative dielectric constant may be a signature of superconductivity.\cite{Reyren07p1196} Here previous observations of superconductivity may be a result of disorder induced effect; with perhaps the degree of disorder giving rise to changes in the superconducting transition temperatures. Another effect could be a route to stabilizing the transient differential negative capacitance previously observed in ferroelectric/dielectric capacitors.\cite{Catalan15p137, Gao14p5814, Khan15p182, Zubko16p524}  

\textbf{\emph{Carrier mobilities.}} To study the effects of intermixing on the mobilities of these materials, we examine the changes in band effective masses $m^*$ as a function of intermixing fraction. Figure~\ref{fig4} lists the average band effective masses of the light-mass bands and the band effective mass of the heavy band for the intermixed systems studied. Here, we see that for both cases there is a decrease in $m^*$ with increased mixing fraction. Similar results were observed for fractionally $\delta$-doped superlattices. In the two-band model, the decrease in $m^*$ was linked to significant enhancements in the overall mobilitiy of the systems. This enhancement was explained as a direct consequence in the loss of the heavy mass band (typically of Ti $d_{\rm xz}+d_{\rm yz}$ character in the $\Gamma$-X/Y directions).\cite{Choi12p4590, Cooper14p6021} It should be pointed out that although these carriers have large effective masses in the $\Gamma$-X/Y  directions they generally give rise to low density high mobility (LDHM) carriers in the $\Gamma$-M direction. This is because of the fact that due to orbital ordering, the electrons in the Ti $d_{\rm xy}$ orbitals are strongly localized and therefore labeled as high density low mobility (HDLM).\cite{Okamoto06p056802,  Kim10p201407, Khalsa13p41302}

%\begin{figure}[b]
%\includegraphics[width=2.5in]{figures/MobileCarriers}
%%action for the abrupt, 1/4, 1/2 and 3/4 intermixed interfaces.}
%\end{figure}

In the fractionally $\delta$-doped study, a strong correlation was observed between the fraction of LDHM carriers and the overall mobility of the system.\cite{Choi12p4590} In short, increases in the mobility of the system, due to increased intermixing, were related with subsequent increases in the relative fraction of LDHM carriers. Table~\ref{tab1} lists the relative fraction of Ti $d_{\rm xz}+d_{\rm yz}$ (LDHM) carriers when going from the abrupt interface to the intermixed interfaces.  Here, we see a substantial increase in the mixing fraction for the 1/4 intermixed interface relative to the abrupt interface. This intermixing fraction decreases as the intermixing fraction increases, suggesting that 1/4 intermixed interfaces may have the highest mobilities due to a larger fraction of mobile electrons. Interestingly, the 1/2 intermixed interface seems to contradict previous results from the fractionally $\delta$-doped interfaces which found the 50 \% intermixng fraction to have the highest observed mobilities. Here, however, the symmetry of the interface needs to be considered. 

\textbf{\emph{Conclusion.}} 
In conclusion, we have studied the effects of interfacial $A$-site intermixing on the charge carrier density and mobilities in 1.5 LAO/8.5 STO superlattices using DFT. Our results demonstrate that the effects of intermixing are strongly localized to the layers near the interface, returning to sharp interface behaviour at only 1 or 2 unit cells away from the interface. Despite this quick turn around we find substantial changes in charge density distributions, polar distortions and electric fields near the interface. Both the charge density distribution (electron count) and polar distortions show large (almost linear with intermixing fraction) decreases near the interface. Surprisingly, the polar distortions also show a change in direction for intermixing fractions of 1/2 or greater. In addition, we find a 32\% enhancement in the magnitude of the electric field away from the interface for the 3/4 intermixing interface, which can be correlated with the shift in peak charge density to one layer away from the AlO$_2$ interface. The combined behaviour of the interfacial and bulk regions are commensurate with the notion that the effect of the interface is simply to shift the potential but not alter its slope. These results point to the possibility of anomalous dielectric responses at the oxide heterostructure interfaces. Such changes may be the reason why some results show superconductivity at oxide heterointerfaces while systems with cleaner interfaces result in the absence of this behavior.\cite{Richter13p528} Another consequence is that this may provide a mechanism for the stabilization of negative differential capacitance at the interface.
Furthermore, within the framework of the two-band model, our results suggest that interfacial intermixing may lead to enhancements in carrier mobilities. This may be driven both through an increase in the fraction of high density carriers as well as a decrease in band effective masses. These effects are similar to those found in previous studies of fractionally $\delta$-doped STO 2DEG systems.

Together, these results suggest that in different samples with different amounts of intermixing there should be a range of carrier concentrations. Indeed, recent work exploring the number of 2DEG charge carriers in a range of LTO/STO superlattices showed that although there was no specific trend in carrier densities with relative fractions of La vs. Sr, there was a significant range of carrier densities. The range of carrier concentrations reported by experiments on these systems seems to be consistent with our predicted reduction in peak carrier concentration of $\sim$25\% when going from the abrupt interface to the 3/4 interface.\cite{Seo07p266801, Kim10p201407} Thus, pointing to the possibility of significant deviations in charge carrier densities due to interfacial intermixing. As such, this study suggests routes by which disorder at an interface can be used to tune (and actually enhance) the properties of carriers at heterostructure interfaces. Furthermore, the larger enthalpy of mixing for intermixing fractions greater than 50\% also point to the possibility of creating these systems with techniques such as chemical vapor deposition, which may have specific consequences for large scale production.

\textbf{\emph{Methodology}} 
All calculations employed the QUANTUM ESPRESSO simulation code (v 5.0.2)\cite{Giannozzi09p395502a} using the local density approximation (LDA) for exchange and correlation and ultrasoft pseudopotentials. The Sr 4$s$4$p$5$s$, Ti 3$p$3$d$4$s$, La 5$s$5$p$5$d$6$s$, Al 3$s$3$p$, and O 2$s$2$p$ electrons were treated as valence electrons. In all cases, 2 $\times$ 2 $\times$ 10 perovskite unit cell superlattices were used to model the system. With a stoichiometry equivalent to 1.5 LAO : 8.5 STO (LaO-AlO$_2$-LaO:TiO$_2$-8 SrTiO$_3$). We used a 500 eV cutoff and a 8 $\times$ 8 $\times$ 1 Monkhorst-Pack $k$-point mesh. The in-plane lattice constants $a$ and $b$ were constrained to the theoretical value of STO (3.863 \AA; obtained from standard LDA without the inclusion of a Hubbard $U$) and the out-of-plane lattice vector $c$ was optimized within the $P4mm$ space group. (N.B. the computed STO lattice constant is in typical LDA agreement with the experimental value of 3.901 \AA). All ionic coordinates were relaxed until all Hellman-Feynman forces were less than 5 meV/\AA. A Hubbard $U$ of 5.0 eV for Ti $d$-states was found to be appropriate and was employed for all superlattice calculations.\cite{Anisimov91p943}  Similar $U$ values were found to be sufficient in previous studies of LaTiO$_3$/STO;\cite{Choi12p4590, Okamoto06p056802, Cooper14p6021, Hamann06p195403} giving a reasonable description of the electronic and structural rearrangements that occur in these materials.

The electric field was computed from the divergence of the $xy$-layer averaged electrostatic potential minus the bare ion potential\cite{Baldereschi88p734}. For the layer averaging we chose an integration window equivalent to roughly one STO perovskite unit cell along the $c$-axis; similar to previous work on 2DEG heterointerfaces \cite{Chen10p85430, Chen10p2881}. The dielectric response was then computed based upon the assumption that the displacement field, $D = 4 \pi P - \epsilon E$, in the superlattice remained constant. We further assume that the Born effective charges $Z$* for La, Sr, and Ti in the superlattice are the same as in bulk LAO  and STO. Our computed $Z$* were 4.3, 2.6, and 5.5 for La, Sr, and Ti, respectively. By taking the derivative of $D$ and setting it to 0 we can then solve for $\epsilon0$ as a function of $z$-axis coordinate in each superlattice. While we anticipate significant deviations in the Born effective charge or polarization in the interfacial 2DEG regions, this approach should give an upper bound on the local dielectric response; thus providing insight into the response of these systems to interfacial intermixing. A similar approach was employed to study the electric deadlayer in polar oxide superlattices\cite{Wang10p094114}. 

\textbf\emph{Acknowledgments.} Initial research was sponsored by the Laboratory Directed Research and Development Program of Oak Ridge National Laboratory, managed by UT-Battelle, LLC, for the US Department of Energy (VRC, HZ, PG, PRCK) and (LZ, HX) The University of Tennessee Science Alliance Joint Directed Research and Development Program and UT/ORNL Joint Institute of Advanced Materials. Followup research by VRC was sponsored by the US Department of Energy (DOE), Office of Science, Basic Energy Sciences (BES), Materials Sciences and Engineering Division through the Office of Science Early Career Research Program. This research used resources of the National Energy Research Scientific Computing Center, which is supported by the Office of Science of the US Department of Energy under Contract No. DE-AC02-05CH11231.

\textbf\emph{Author Contributions.} VRC conceived the project and performed the calculations. H.Z. performed analysis of the electrostatic potential. P.G., H.X. and L.Z. contributed to data analysis particularly with regards to the dielectric response and connections to negative capacitance. P.R.C.K. contributed to data analysis and writing of the manuscript.

\bibliography{citations}

\begin{thebibliography}{36}%
\makeatletter
\providecommand \@ifxundefined [1]{%
 \@ifx{#1\undefined}
}%
\providecommand \@ifnum [1]{%
 \ifnum #1\expandafter \@firstoftwo
 \else \expandafter \@secondoftwo
 \fi
}%
\providecommand \@ifx [1]{%
 \ifx #1\expandafter \@firstoftwo
 \else \expandafter \@secondoftwo
 \fi
}%
\providecommand \natexlab [1]{#1}%
\providecommand \enquote  [1]{``#1''}%
\providecommand \bibnamefont  [1]{#1}%
\providecommand \bibfnamefont [1]{#1}%
\providecommand \citenamefont [1]{#1}%
\providecommand \href@noop [0]{\@secondoftwo}%
\providecommand \href [0]{\begingroup \@sanitize@url \@href}%
\providecommand \@href[1]{\@@startlink{#1}\@@href}%
\providecommand \@@href[1]{\endgroup#1\@@endlink}%
\providecommand \@sanitize@url [0]{\catcode `\\12\catcode `\$12\catcode
  `\&12\catcode `\#12\catcode `\^12\catcode `\_12\catcode `\%12\relax}%
\providecommand \@@startlink[1]{}%
\providecommand \@@endlink[0]{}%
\providecommand \url  [0]{\begingroup\@sanitize@url \@url }%
\providecommand \@url [1]{\endgroup\@href {#1}{\urlprefix }}%
\providecommand \urlprefix  [0]{URL }%
\providecommand \Eprint [0]{\href }%
\providecommand \doibase [0]{http://dx.doi.org/}%
\providecommand \selectlanguage [0]{\@gobble}%
\providecommand \bibinfo  [0]{\@secondoftwo}%
\providecommand \bibfield  [0]{\@secondoftwo}%
\providecommand \translation [1]{[#1]}%
\providecommand \BibitemOpen [0]{}%
\providecommand \bibitemStop [0]{}%
\providecommand \bibitemNoStop [0]{.\EOS\space}%
\providecommand \EOS [0]{\spacefactor3000\relax}%
\providecommand \BibitemShut  [1]{\csname bibitem#1\endcsname}%
\let\auto@bib@innerbib\@empty
%</preamble>
\bibitem [{\citenamefont {Okamoto}\ and\ \citenamefont
  {Millis}(2004)}]{Okamoto04p630}%
  \BibitemOpen
  \bibfield  {author} {\bibinfo {author} {\bibfnamefont {S.}~\bibnamefont
  {Okamoto}}\ and\ \bibinfo {author} {\bibfnamefont {A.~J.}\ \bibnamefont
  {Millis}},\ }\href@noop {} {\bibfield  {journal} {\bibinfo  {journal}
  {Nature}\ }\textbf {\bibinfo {volume} {428}},\ \bibinfo {pages} {630}
  (\bibinfo {year} {2004})}\BibitemShut {NoStop}%
\bibitem [{\citenamefont {Ohtomo}\ \emph {et~al.}(2002)\citenamefont {Ohtomo},
  \citenamefont {Muller}, \citenamefont {Grazul},\ and\ \citenamefont
  {Hwang}}]{Ohtomo02p378}%
  \BibitemOpen
  \bibfield  {author} {\bibinfo {author} {\bibfnamefont {A.}~\bibnamefont
  {Ohtomo}}, \bibinfo {author} {\bibfnamefont {D.~A.}\ \bibnamefont {Muller}},
  \bibinfo {author} {\bibfnamefont {J.~L.}\ \bibnamefont {Grazul}}, \ and\
  \bibinfo {author} {\bibfnamefont {H.~Y.}\ \bibnamefont {Hwang}},\ }\href@noop
  {} {\bibfield  {journal} {\bibinfo  {journal} {Nature}\ }\textbf {\bibinfo
  {volume} {419}},\ \bibinfo {pages} {378} (\bibinfo {year}
  {2002})}\BibitemShut {NoStop}%
\bibitem [{\citenamefont {Ohtomo}\ and\ \citenamefont
  {Hwang}(2004)}]{Ohtomo04p423}%
  \BibitemOpen
  \bibfield  {author} {\bibinfo {author} {\bibfnamefont {A.}~\bibnamefont
  {Ohtomo}}\ and\ \bibinfo {author} {\bibfnamefont {H.~Y.}\ \bibnamefont
  {Hwang}},\ }\href@noop {} {\bibfield  {journal} {\bibinfo  {journal}
  {Nature}\ }\textbf {\bibinfo {volume} {427}},\ \bibinfo {pages} {423}
  (\bibinfo {year} {2004})}\BibitemShut {NoStop}%
\bibitem [{\citenamefont {Shibuya}\ \emph {et~al.}(2004)\citenamefont
  {Shibuya}, \citenamefont {Ohnishi}, \citenamefont {Kawasaki}, \citenamefont
  {Koinuma},\ and\ \citenamefont {Lippmaa}}]{Shibuya04pL1178}%
  \BibitemOpen
  \bibfield  {author} {\bibinfo {author} {\bibfnamefont {K.}~\bibnamefont
  {Shibuya}}, \bibinfo {author} {\bibfnamefont {T.}~\bibnamefont {Ohnishi}},
  \bibinfo {author} {\bibfnamefont {M.}~\bibnamefont {Kawasaki}}, \bibinfo
  {author} {\bibfnamefont {H.}~\bibnamefont {Koinuma}}, \ and\ \bibinfo
  {author} {\bibfnamefont {M.}~\bibnamefont {Lippmaa}},\ }\href@noop {}
  {\bibfield  {journal} {\bibinfo  {journal} {Jap. J. App. Phys.}\ }\textbf
  {\bibinfo {volume} {43}},\ \bibinfo {pages} {L1178} (\bibinfo {year}
  {2004})}\BibitemShut {NoStop}%
\bibitem [{\citenamefont {Hwang}\ \emph {et~al.}(2012)\citenamefont {Hwang},
  \citenamefont {Iwasa}, \citenamefont {Kawasaki}, \citenamefont {Keimer},
  \citenamefont {Nagaosa},\ and\ \citenamefont {Tokura}}]{Hwang12p103}%
  \BibitemOpen
  \bibfield  {author} {\bibinfo {author} {\bibfnamefont {H.~Y.}\ \bibnamefont
  {Hwang}}, \bibinfo {author} {\bibfnamefont {Y.}~\bibnamefont {Iwasa}},
  \bibinfo {author} {\bibfnamefont {M.}~\bibnamefont {Kawasaki}}, \bibinfo
  {author} {\bibfnamefont {B.}~\bibnamefont {Keimer}}, \bibinfo {author}
  {\bibfnamefont {N.}~\bibnamefont {Nagaosa}}, \ and\ \bibinfo {author}
  {\bibfnamefont {Y.}~\bibnamefont {Tokura}},\ }\href@noop {} {\bibfield
  {journal} {\bibinfo  {journal} {Nat. Mater.}\ }\textbf {\bibinfo {volume}
  {11}},\ \bibinfo {pages} {103} (\bibinfo {year} {2012})}\BibitemShut
  {NoStop}%
\bibitem [{\citenamefont {Zubko}\ \emph {et~al.}(2011)\citenamefont {Zubko},
  \citenamefont {Gariglio}, \citenamefont {Gabay}, \citenamefont {Ghosez},\
  and\ \citenamefont {Triscone}}]{Zubko11p141}%
  \BibitemOpen
  \bibfield  {author} {\bibinfo {author} {\bibfnamefont {P.}~\bibnamefont
  {Zubko}}, \bibinfo {author} {\bibfnamefont {S.}~\bibnamefont {Gariglio}},
  \bibinfo {author} {\bibfnamefont {M.}~\bibnamefont {Gabay}}, \bibinfo
  {author} {\bibfnamefont {P.}~\bibnamefont {Ghosez}}, \ and\ \bibinfo {author}
  {\bibfnamefont {J.-M.}\ \bibnamefont {Triscone}},\ }\href {\doibase
  10.1146/annurev-conmatphys-062910-140445} {\bibfield  {journal} {\bibinfo
  {journal} {Annu. Rev. Condens. Matter Phys.}\ }\textbf {\bibinfo {volume}
  {2}},\ \bibinfo {pages} {141} (\bibinfo {year} {2011})}\BibitemShut {NoStop}%
\bibitem [{\citenamefont {Mannhart}\ and\ \citenamefont
  {Schlom}(2010)}]{Mannhart10p1607}%
  \BibitemOpen
  \bibfield  {author} {\bibinfo {author} {\bibfnamefont {J.}~\bibnamefont
  {Mannhart}}\ and\ \bibinfo {author} {\bibfnamefont {D.~G.}\ \bibnamefont
  {Schlom}},\ }\href {\doibase 10.1126/science.1181862} {\bibfield  {journal}
  {\bibinfo  {journal} {Science}\ }\textbf {\bibinfo {volume} {327}},\ \bibinfo
  {pages} {1607} (\bibinfo {year} {2010})}\BibitemShut {NoStop}%
\bibitem [{\citenamefont {Reyren}\ \emph {et~al.}(2007)\citenamefont {Reyren},
  \citenamefont {Thiel}, \citenamefont {Caviglia}, \citenamefont {Kourkoutis},
  \citenamefont {Hammerl}, \citenamefont {Richter}, \citenamefont {Schneider},
  \citenamefont {Kopp}, \citenamefont {Rüetschi}, \citenamefont {Jaccard},
  \citenamefont {Gabay}, \citenamefont {Muller}, \citenamefont {Triscone},\
  and\ \citenamefont {Mannhart}}]{Reyren07p1196}%
  \BibitemOpen
  \bibfield  {author} {\bibinfo {author} {\bibfnamefont {N.}~\bibnamefont
  {Reyren}}, \bibinfo {author} {\bibfnamefont {S.}~\bibnamefont {Thiel}},
  \bibinfo {author} {\bibfnamefont {A.~D.}\ \bibnamefont {Caviglia}}, \bibinfo
  {author} {\bibfnamefont {L.~F.}\ \bibnamefont {Kourkoutis}}, \bibinfo
  {author} {\bibfnamefont {G.}~\bibnamefont {Hammerl}}, \bibinfo {author}
  {\bibfnamefont {C.}~\bibnamefont {Richter}}, \bibinfo {author} {\bibfnamefont
  {C.~W.}\ \bibnamefont {Schneider}}, \bibinfo {author} {\bibfnamefont
  {T.}~\bibnamefont {Kopp}}, \bibinfo {author} {\bibfnamefont {A.-S.}\
  \bibnamefont {Rüetschi}}, \bibinfo {author} {\bibfnamefont {D.}~\bibnamefont
  {Jaccard}}, \bibinfo {author} {\bibfnamefont {M.}~\bibnamefont {Gabay}},
  \bibinfo {author} {\bibfnamefont {D.~A.}\ \bibnamefont {Muller}}, \bibinfo
  {author} {\bibfnamefont {J.-M.}\ \bibnamefont {Triscone}}, \ and\ \bibinfo
  {author} {\bibfnamefont {J.}~\bibnamefont {Mannhart}},\ }\href {\doibase
  10.1126/science.1146006} {\bibfield  {journal} {\bibinfo  {journal}
  {Science}\ }\textbf {\bibinfo {volume} {317}},\ \bibinfo {pages} {1196}
  (\bibinfo {year} {2007})}\BibitemShut {NoStop}%
\bibitem [{\citenamefont {Brinkman}\ \emph {et~al.}(2007)\citenamefont
  {Brinkman}, \citenamefont {Huijben}, \citenamefont {van Zalk}, \citenamefont
  {Huijben}, \citenamefont {Zeitler}, \citenamefont {Maan}, \citenamefont
  {van~der Wiel}, \citenamefont {Rijnders}, \citenamefont {Blank},\ and\
  \citenamefont {Hilgenkamp}}]{Brinkman07p493}%
  \BibitemOpen
  \bibfield  {author} {\bibinfo {author} {\bibfnamefont {A.}~\bibnamefont
  {Brinkman}}, \bibinfo {author} {\bibfnamefont {M.}~\bibnamefont {Huijben}},
  \bibinfo {author} {\bibfnamefont {M.}~\bibnamefont {van Zalk}}, \bibinfo
  {author} {\bibfnamefont {J.}~\bibnamefont {Huijben}}, \bibinfo {author}
  {\bibfnamefont {U.}~\bibnamefont {Zeitler}}, \bibinfo {author} {\bibfnamefont
  {J.~C.}\ \bibnamefont {Maan}}, \bibinfo {author} {\bibfnamefont {W.~G.}\
  \bibnamefont {van~der Wiel}}, \bibinfo {author} {\bibfnamefont
  {G.}~\bibnamefont {Rijnders}}, \bibinfo {author} {\bibfnamefont {D.~H.~A.}\
  \bibnamefont {Blank}}, \ and\ \bibinfo {author} {\bibfnamefont
  {H.}~\bibnamefont {Hilgenkamp}},\ }\href@noop {} {\bibfield  {journal}
  {\bibinfo  {journal} {Nat. Mater.}\ }\textbf {\bibinfo {volume} {6}},\
  \bibinfo {pages} {493} (\bibinfo {year} {2007})}\BibitemShut {NoStop}%
\bibitem [{\citenamefont {Glavic}\ \emph {et~al.}(2016)\citenamefont {Glavic},
  \citenamefont {Dixit}, \citenamefont {Cooper},\ and\ \citenamefont
  {Aczel}}]{Glavic16p140413}%
  \BibitemOpen
  \bibfield  {author} {\bibinfo {author} {\bibfnamefont {A.}~\bibnamefont
  {Glavic}}, \bibinfo {author} {\bibfnamefont {H.}~\bibnamefont {Dixit}},
  \bibinfo {author} {\bibfnamefont {V.~R.}\ \bibnamefont {Cooper}}, \ and\
  \bibinfo {author} {\bibfnamefont {A.~A.}\ \bibnamefont {Aczel}},\ }\href
  {\doibase 10.1103/PhysRevB.93.140413} {\bibfield  {journal} {\bibinfo
  {journal} {Phys. Rev. B}\ }\textbf {\bibinfo {volume} {93}},\ \bibinfo
  {pages} {140413} (\bibinfo {year} {2016})}\BibitemShut {NoStop}%
\bibitem [{\citenamefont {Nakagawa}\ \emph {et~al.}(2006)\citenamefont
  {Nakagawa}, \citenamefont {Hwang},\ and\ \citenamefont
  {Muller}}]{Nakagawa06p204}%
  \BibitemOpen
  \bibfield  {author} {\bibinfo {author} {\bibfnamefont {N.}~\bibnamefont
  {Nakagawa}}, \bibinfo {author} {\bibfnamefont {H.~Y.}\ \bibnamefont {Hwang}},
  \ and\ \bibinfo {author} {\bibfnamefont {D.~A.}\ \bibnamefont {Muller}},\
  }\href@noop {} {\bibfield  {journal} {\bibinfo  {journal} {Nat. Mater.}\
  }\textbf {\bibinfo {volume} {5}},\ \bibinfo {pages} {204} (\bibinfo {year}
  {2006})}\BibitemShut {NoStop}%
\bibitem [{\citenamefont {Cooper}\ \emph {et~al.}(2007)\citenamefont {Cooper},
  \citenamefont {Johnston},\ and\ \citenamefont {Rabe}}]{Cooper07p020103R}%
  \BibitemOpen
  \bibfield  {author} {\bibinfo {author} {\bibfnamefont {V.~R.}\ \bibnamefont
  {Cooper}}, \bibinfo {author} {\bibfnamefont {K.}~\bibnamefont {Johnston}}, \
  and\ \bibinfo {author} {\bibfnamefont {K.~M.}\ \bibnamefont {Rabe}},\
  }\href@noop {} {\bibfield  {journal} {\bibinfo  {journal} {Phys. Rev. B~}\
  }\textbf {\bibinfo {volume} {76}},\ \bibinfo {pages} {020103(R)} (\bibinfo
  {year} {2007})}\BibitemShut {NoStop}%
\bibitem [{\citenamefont {Chen}\ and\ \citenamefont
  {Millis}(2016)}]{Chen16p104111}%
  \BibitemOpen
  \bibfield  {author} {\bibinfo {author} {\bibfnamefont {H.}~\bibnamefont
  {Chen}}\ and\ \bibinfo {author} {\bibfnamefont {A.}~\bibnamefont {Millis}},\
  }\href {\doibase 10.1103/PhysRevB.93.104111} {\bibfield  {journal} {\bibinfo
  {journal} {Phys. Rev. B}\ }\textbf {\bibinfo {volume} {93}},\ \bibinfo
  {pages} {104111} (\bibinfo {year} {2016})}\BibitemShut {NoStop}%
\bibitem [{\citenamefont {Choi}\ \emph {et~al.}(2012)\citenamefont {Choi},
  \citenamefont {Lee}, \citenamefont {Cooper},\ and\ \citenamefont
  {Lee}}]{Choi12p4590}%
  \BibitemOpen
  \bibfield  {author} {\bibinfo {author} {\bibfnamefont {W.~S.}\ \bibnamefont
  {Choi}}, \bibinfo {author} {\bibfnamefont {S.}~\bibnamefont {Lee}}, \bibinfo
  {author} {\bibfnamefont {V.~R.}\ \bibnamefont {Cooper}}, \ and\ \bibinfo
  {author} {\bibfnamefont {H.~N.}\ \bibnamefont {Lee}},\ }\href@noop {}
  {\bibfield  {journal} {\bibinfo  {journal} {Nano Lett.}\ }\textbf {\bibinfo
  {volume} {12}},\ \bibinfo {pages} {4590} (\bibinfo {year}
  {2012})}\BibitemShut {NoStop}%
\bibitem [{\citenamefont {Cooper}(2012)}]{Cooper12p235109}%
  \BibitemOpen
  \bibfield  {author} {\bibinfo {author} {\bibfnamefont {V.~R.}\ \bibnamefont
  {Cooper}},\ }\href@noop {} {\bibfield  {journal} {\bibinfo  {journal} {Phys.
  Rev. B}\ }\textbf {\bibinfo {volume} {85}},\ \bibinfo {pages} {235109}
  (\bibinfo {year} {2012})}\BibitemShut {NoStop}%
\bibitem [{\citenamefont {Shen}\ \emph {et~al.}(2015)\citenamefont {Shen},
  \citenamefont {Wang}, \citenamefont {Zhou}, \citenamefont {Jiang},
  \citenamefont {Hou},\ and\ \citenamefont {Fei}}]{Shen15p74}%
  \BibitemOpen
  \bibfield  {author} {\bibinfo {author} {\bibfnamefont {Y.}~\bibnamefont
  {Shen}}, \bibinfo {author} {\bibfnamefont {W.}~\bibnamefont {Wang}}, \bibinfo
  {author} {\bibfnamefont {Z.}~\bibnamefont {Zhou}}, \bibinfo {author}
  {\bibfnamefont {Y.}~\bibnamefont {Jiang}}, \bibinfo {author} {\bibfnamefont
  {C.}~\bibnamefont {Hou}}, \ and\ \bibinfo {author} {\bibfnamefont
  {W.}~\bibnamefont {Fei}},\ }\href {\doibase 10.1007/s10853-014-8567-7}
  {\bibfield  {journal} {\bibinfo  {journal} {J. Mater. Sci.}\ }\textbf
  {\bibinfo {volume} {50}},\ \bibinfo {pages} {74} (\bibinfo {year}
  {2015})}\BibitemShut {NoStop}%
\bibitem [{\citenamefont {Zou}\ \emph {et~al.}(2015)\citenamefont {Zou},
  \citenamefont {Ismail-Beigi}, \citenamefont {Kisslinger}, \citenamefont
  {Shen}, \citenamefont {Su}, \citenamefont {Walker},\ and\ \citenamefont
  {Ahn}}]{Zou15p36104}%
  \BibitemOpen
  \bibfield  {author} {\bibinfo {author} {\bibfnamefont {K.}~\bibnamefont
  {Zou}}, \bibinfo {author} {\bibfnamefont {S.}~\bibnamefont {Ismail-Beigi}},
  \bibinfo {author} {\bibfnamefont {K.}~\bibnamefont {Kisslinger}}, \bibinfo
  {author} {\bibfnamefont {X.}~\bibnamefont {Shen}}, \bibinfo {author}
  {\bibfnamefont {D.}~\bibnamefont {Su}}, \bibinfo {author} {\bibfnamefont
  {F.~J.}\ \bibnamefont {Walker}}, \ and\ \bibinfo {author} {\bibfnamefont
  {C.~H.}\ \bibnamefont {Ahn}},\ }\href {\doibase
  http://dx.doi.org/10.1063/1.4914310} {\bibfield  {journal} {\bibinfo
  {journal} {APL Mater.}\ }\textbf {\bibinfo {volume} {3}},\ \bibinfo {eid}
  {036104} (\bibinfo {year} {2015})}\BibitemShut {NoStop}%
\bibitem [{\citenamefont {Seo}\ \emph {et~al.}(2007)\citenamefont {Seo},
  \citenamefont {Choi}, \citenamefont {Lee}, \citenamefont {Yu}, \citenamefont
  {Kim}, \citenamefont {Bernhard},\ and\ \citenamefont {Noh}}]{Seo07p266801}%
  \BibitemOpen
  \bibfield  {author} {\bibinfo {author} {\bibfnamefont {S.~S.~A.}\
  \bibnamefont {Seo}}, \bibinfo {author} {\bibfnamefont {W.~S.}\ \bibnamefont
  {Choi}}, \bibinfo {author} {\bibfnamefont {H.~N.}\ \bibnamefont {Lee}},
  \bibinfo {author} {\bibfnamefont {L.}~\bibnamefont {Yu}}, \bibinfo {author}
  {\bibfnamefont {K.~W.}\ \bibnamefont {Kim}}, \bibinfo {author} {\bibfnamefont
  {C.}~\bibnamefont {Bernhard}}, \ and\ \bibinfo {author} {\bibfnamefont
  {T.~W.}\ \bibnamefont {Noh}},\ }\href@noop {} {\bibfield  {journal} {\bibinfo
   {journal} {Phys. Rev. Lett.}\ }\textbf {\bibinfo {volume} {99}},\ \bibinfo
  {pages} {266801} (\bibinfo {year} {2007})}\BibitemShut {NoStop}%
\bibitem [{\citenamefont {Catalan}\ \emph {et~al.}(2015)\citenamefont
  {Catalan}, \citenamefont {Jim\'{e}nez},\ and\ \citenamefont
  {Gruverman}}]{Catalan15p137}%
  \BibitemOpen
  \bibfield  {author} {\bibinfo {author} {\bibfnamefont {G.}~\bibnamefont
  {Catalan}}, \bibinfo {author} {\bibfnamefont {D.}~\bibnamefont
  {Jim\'{e}nez}}, \ and\ \bibinfo {author} {\bibfnamefont {A.}~\bibnamefont
  {Gruverman}},\ }\href@noop {} {\bibfield  {journal} {\bibinfo  {journal}
  {Nat. Mater.}\ }\textbf {\bibinfo {volume} {14}},\ \bibinfo {pages} {137}
  (\bibinfo {year} {2015})}\BibitemShut {NoStop}%
\bibitem [{\citenamefont {Gao}\ \emph {et~al.}(2014)\citenamefont {Gao},
  \citenamefont {Khan}, \citenamefont {Marti}, \citenamefont {Nelson},
  \citenamefont {Serrao}, \citenamefont {Ravichandran}, \citenamefont
  {Ramesh},\ and\ \citenamefont {Salahuddin}}]{Gao14p5814}%
  \BibitemOpen
  \bibfield  {author} {\bibinfo {author} {\bibfnamefont {W.}~\bibnamefont
  {Gao}}, \bibinfo {author} {\bibfnamefont {A.}~\bibnamefont {Khan}}, \bibinfo
  {author} {\bibfnamefont {X.}~\bibnamefont {Marti}}, \bibinfo {author}
  {\bibfnamefont {C.}~\bibnamefont {Nelson}}, \bibinfo {author} {\bibfnamefont
  {C.}~\bibnamefont {Serrao}}, \bibinfo {author} {\bibfnamefont
  {J.}~\bibnamefont {Ravichandran}}, \bibinfo {author} {\bibfnamefont
  {R.}~\bibnamefont {Ramesh}}, \ and\ \bibinfo {author} {\bibfnamefont
  {S.}~\bibnamefont {Salahuddin}},\ }\href {\doibase 10.1021/nl502691u}
  {\bibfield  {journal} {\bibinfo  {journal} {Nano Lett.}\ }\textbf {\bibinfo
  {volume} {14}},\ \bibinfo {pages} {5814} (\bibinfo {year}
  {2014})}\BibitemShut {NoStop}%
\bibitem [{\citenamefont {Khan}\ \emph {et~al.}(2015)\citenamefont {Khan},
  \citenamefont {Chatterjee}, \citenamefont {Wang}, \citenamefont {Drapcho},
  \citenamefont {You}, \citenamefont {Serrao}, \citenamefont {Bakaul},\ and\
  \citenamefont {Salahuddin}}]{Khan15p182}%
  \BibitemOpen
  \bibfield  {author} {\bibinfo {author} {\bibfnamefont {A.~I.}\ \bibnamefont
  {Khan}}, \bibinfo {author} {\bibfnamefont {K.}~\bibnamefont {Chatterjee}},
  \bibinfo {author} {\bibfnamefont {B.}~\bibnamefont {Wang}}, \bibinfo {author}
  {\bibfnamefont {S.}~\bibnamefont {Drapcho}}, \bibinfo {author} {\bibfnamefont
  {L.}~\bibnamefont {You}}, \bibinfo {author} {\bibfnamefont {C.}~\bibnamefont
  {Serrao}}, \bibinfo {author} {\bibfnamefont {S.~R.}\ \bibnamefont {Bakaul}},
  \ and\ \bibinfo {author} {\bibfnamefont {R.~R.~S.}\ \bibnamefont
  {Salahuddin}},\ }\href@noop {} {\bibfield  {journal} {\bibinfo  {journal}
  {Nat. Mater.}\ }\textbf {\bibinfo {volume} {14}},\ \bibinfo {pages} {182}
  (\bibinfo {year} {2015})}\BibitemShut {NoStop}%
\bibitem [{\citenamefont {Zubko}\ \emph {et~al.}(2016)\citenamefont {Zubko},
  \citenamefont {Wojde{\l}}, \citenamefont {Hadjimichael}, \citenamefont
  {Fernandez-Pena}, \citenamefont {Sen{\'{e}}}, \citenamefont
  {Luk{\textquoteright}yanchuk}, \citenamefont {Triscone},\ and\ \citenamefont
  {{\'{I}}{\~{n}}iguez}}]{Zubko16p524}%
  \BibitemOpen
  \bibfield  {author} {\bibinfo {author} {\bibfnamefont {P.}~\bibnamefont
  {Zubko}}, \bibinfo {author} {\bibfnamefont {J.~C.}\ \bibnamefont
  {Wojde{\l}}}, \bibinfo {author} {\bibfnamefont {M.}~\bibnamefont
  {Hadjimichael}}, \bibinfo {author} {\bibfnamefont {S.}~\bibnamefont
  {Fernandez-Pena}}, \bibinfo {author} {\bibfnamefont {A.}~\bibnamefont
  {Sen{\'{e}}}}, \bibinfo {author} {\bibfnamefont {I.}~\bibnamefont
  {Luk{\textquoteright}yanchuk}}, \bibinfo {author} {\bibfnamefont {J.-M.}\
  \bibnamefont {Triscone}}, \ and\ \bibinfo {author} {\bibfnamefont
  {J.}~\bibnamefont {{\'{I}}{\~{n}}iguez}},\ }\href@noop {} {\bibfield
  {journal} {\bibinfo  {journal} {Nature}\ }\textbf {\bibinfo {volume} {534}},\
  \bibinfo {pages} {524} (\bibinfo {year} {2016})}\BibitemShut {NoStop}%
\bibitem [{\citenamefont {Cantoni}\ \emph {et~al.}(2012)\citenamefont
  {Cantoni}, \citenamefont {Gazquez}, \citenamefont {Granozio}, \citenamefont
  {Oxley}, \citenamefont {Varela}, \citenamefont {Lupini}, \citenamefont
  {Pennycook}, \citenamefont {Aruta}, \citenamefont {di~Uccio}, \citenamefont
  {Perna},\ and\ \citenamefont {Maccariello}}]{Cantoni12p3952}%
  \BibitemOpen
  \bibfield  {author} {\bibinfo {author} {\bibfnamefont {C.}~\bibnamefont
  {Cantoni}}, \bibinfo {author} {\bibfnamefont {J.}~\bibnamefont {Gazquez}},
  \bibinfo {author} {\bibfnamefont {F.~M.}\ \bibnamefont {Granozio}}, \bibinfo
  {author} {\bibfnamefont {M.~P.}\ \bibnamefont {Oxley}}, \bibinfo {author}
  {\bibfnamefont {M.}~\bibnamefont {Varela}}, \bibinfo {author} {\bibfnamefont
  {A.~R.}\ \bibnamefont {Lupini}}, \bibinfo {author} {\bibfnamefont {S.~J.}\
  \bibnamefont {Pennycook}}, \bibinfo {author} {\bibfnamefont {C.}~\bibnamefont
  {Aruta}}, \bibinfo {author} {\bibfnamefont {U.~S.}\ \bibnamefont {di~Uccio}},
  \bibinfo {author} {\bibfnamefont {P.}~\bibnamefont {Perna}}, \ and\ \bibinfo
  {author} {\bibfnamefont {D.}~\bibnamefont {Maccariello}},\ }\href {\doibase
  {10.1002/adma.201200667}} {\bibfield  {journal} {\bibinfo  {journal} {Adv.
  Mater.}\ }\textbf {\bibinfo {volume} {24}},\ \bibinfo {pages} {3952}
  (\bibinfo {year} {2012})}\BibitemShut {NoStop}%
\bibitem [{\citenamefont {Bristowe}\ \emph {et~al.}(2014)\citenamefont
  {Bristowe}, \citenamefont {Ghosez}, \citenamefont {Littlewood},\ and\
  \citenamefont {Artacho}}]{Bristowe14p143201}%
  \BibitemOpen
  \bibfield  {author} {\bibinfo {author} {\bibfnamefont {N.~C.}\ \bibnamefont
  {Bristowe}}, \bibinfo {author} {\bibfnamefont {P.}~\bibnamefont {Ghosez}},
  \bibinfo {author} {\bibfnamefont {P.~B.}\ \bibnamefont {Littlewood}}, \ and\
  \bibinfo {author} {\bibfnamefont {E.}~\bibnamefont {Artacho}},\ }\href
  {http://stacks.iop.org/0953-8984/26/i=14/a=143201} {\bibfield  {journal}
  {\bibinfo  {journal} {J. Phys.: Condens. Matter}\ }\textbf {\bibinfo {volume}
  {26}},\ \bibinfo {pages} {143201} (\bibinfo {year} {2014})}\BibitemShut
  {NoStop}%
\bibitem [{\citenamefont {Cooper}\ \emph {et~al.}(2014)\citenamefont {Cooper},
  \citenamefont {Seo}, \citenamefont {Lee}, \citenamefont {Kim}, \citenamefont
  {Choi}, \citenamefont {Okamoto},\ and\ \citenamefont {Lee}}]{Cooper14p6021}%
  \BibitemOpen
  \bibfield  {author} {\bibinfo {author} {\bibfnamefont {V.~R.}\ \bibnamefont
  {Cooper}}, \bibinfo {author} {\bibfnamefont {S.~S.~A.}\ \bibnamefont {Seo}},
  \bibinfo {author} {\bibfnamefont {S.}~\bibnamefont {Lee}}, \bibinfo {author}
  {\bibfnamefont {J.~S.}\ \bibnamefont {Kim}}, \bibinfo {author} {\bibfnamefont
  {W.~S.}\ \bibnamefont {Choi}}, \bibinfo {author} {\bibfnamefont
  {S.}~\bibnamefont {Okamoto}}, \ and\ \bibinfo {author} {\bibfnamefont
  {H.~N.}\ \bibnamefont {Lee}},\ }\href@noop {} {\bibfield  {journal} {\bibinfo
   {journal} {Sci. Rep.}\ }\textbf {\bibinfo {volume} {4}},\ \bibinfo {pages}
  {6021} (\bibinfo {year} {2014})}\BibitemShut {NoStop}%
\bibitem [{\citenamefont {Okamoto}\ \emph {et~al.}(2006)\citenamefont
  {Okamoto}, \citenamefont {Millis},\ and\ \citenamefont
  {Spaldin}}]{Okamoto06p056802}%
  \BibitemOpen
  \bibfield  {author} {\bibinfo {author} {\bibfnamefont {S.}~\bibnamefont
  {Okamoto}}, \bibinfo {author} {\bibfnamefont {A.~J.}\ \bibnamefont {Millis}},
  \ and\ \bibinfo {author} {\bibfnamefont {N.~A.}\ \bibnamefont {Spaldin}},\
  }\href {\doibase 10.1103/PhysRevLett.97.056802} {\bibfield  {journal}
  {\bibinfo  {journal} {Phys. Rev. Lett.}\ }\textbf {\bibinfo {volume} {97}},\
  \bibinfo {pages} {056802} (\bibinfo {year} {2006})}\BibitemShut {NoStop}%
\bibitem [{\citenamefont {Kim}\ \emph {et~al.}(2010)\citenamefont {Kim},
  \citenamefont {Seo}, \citenamefont {Chisholm}, \citenamefont {Kremer},
  \citenamefont {Habermeier}, \citenamefont {Keimer},\ and\ \citenamefont
  {Lee}}]{Kim10p201407}%
  \BibitemOpen
  \bibfield  {author} {\bibinfo {author} {\bibfnamefont {J.~S.}\ \bibnamefont
  {Kim}}, \bibinfo {author} {\bibfnamefont {S.~S.~A.}\ \bibnamefont {Seo}},
  \bibinfo {author} {\bibfnamefont {M.~F.}\ \bibnamefont {Chisholm}}, \bibinfo
  {author} {\bibfnamefont {R.~K.}\ \bibnamefont {Kremer}}, \bibinfo {author}
  {\bibfnamefont {H.-U.}\ \bibnamefont {Habermeier}}, \bibinfo {author}
  {\bibfnamefont {B.}~\bibnamefont {Keimer}}, \ and\ \bibinfo {author}
  {\bibfnamefont {H.~N.}\ \bibnamefont {Lee}},\ }\href@noop {} {\bibfield
  {journal} {\bibinfo  {journal} {Phys. Rev. B}\ }\textbf {\bibinfo {volume}
  {82}},\ \bibinfo {pages} {201407(R)} (\bibinfo {year} {2010})}\BibitemShut
  {NoStop}%
\bibitem [{\citenamefont {Khalsa}\ \emph {et~al.}(2013)\citenamefont {Khalsa},
  \citenamefont {Lee},\ and\ \citenamefont {MacDonald}}]{Khalsa13p41302}%
  \BibitemOpen
  \bibfield  {author} {\bibinfo {author} {\bibfnamefont {G.}~\bibnamefont
  {Khalsa}}, \bibinfo {author} {\bibfnamefont {B.}~\bibnamefont {Lee}}, \ and\
  \bibinfo {author} {\bibfnamefont {A.~H.}\ \bibnamefont {MacDonald}},\ }\href
  {\doibase 10.1103/PhysRevB.88.041302} {\bibfield  {journal} {\bibinfo
  {journal} {Phys. Rev. B}\ }\textbf {\bibinfo {volume} {88}},\ \bibinfo
  {pages} {041302} (\bibinfo {year} {2013})}\BibitemShut {NoStop}%
\bibitem [{\citenamefont {Richter}\ \emph {et~al.}(2013)\citenamefont
  {Richter}, \citenamefont {Boschker}, \citenamefont {Dietsche}, \citenamefont
  {Fillis-Tsirakis}, \citenamefont {Jany}, \citenamefont {Loder}, \citenamefont
  {Kourkoutis}, \citenamefont {Muller}, \citenamefont {Kirtley}, \citenamefont
  {Schneider},\ and\ \citenamefont {Mannhart}}]{Richter13p528}%
  \BibitemOpen
  \bibfield  {author} {\bibinfo {author} {\bibfnamefont {C.}~\bibnamefont
  {Richter}}, \bibinfo {author} {\bibfnamefont {H.}~\bibnamefont {Boschker}},
  \bibinfo {author} {\bibfnamefont {W.}~\bibnamefont {Dietsche}}, \bibinfo
  {author} {\bibfnamefont {E.}~\bibnamefont {Fillis-Tsirakis}}, \bibinfo
  {author} {\bibfnamefont {R.}~\bibnamefont {Jany}}, \bibinfo {author}
  {\bibfnamefont {F.}~\bibnamefont {Loder}}, \bibinfo {author} {\bibfnamefont
  {L.~F.}\ \bibnamefont {Kourkoutis}}, \bibinfo {author} {\bibfnamefont
  {D.~A.}\ \bibnamefont {Muller}}, \bibinfo {author} {\bibfnamefont {J.~R.}\
  \bibnamefont {Kirtley}}, \bibinfo {author} {\bibfnamefont {C.~W.}\
  \bibnamefont {Schneider}}, \ and\ \bibinfo {author} {\bibfnamefont
  {J.}~\bibnamefont {Mannhart}},\ }\href@noop {} {\bibfield  {journal}
  {\bibinfo  {journal} {Nature}\ }\textbf {\bibinfo {volume} {502}},\ \bibinfo
  {pages} {528} (\bibinfo {year} {2013})}\BibitemShut {NoStop}%
\bibitem [{\citenamefont {Giannozzi}\ \emph {et~al.}(2009)\citenamefont
  {Giannozzi}, \citenamefont {Baroni1}, \citenamefont {Bonini}, \citenamefont
  {Calandra}, \citenamefont {Car}, \citenamefont {Cavazzoni}, \citenamefont
  {Ceresoli}, \citenamefont {Chiarotti}, \citenamefont {Cococcioni},
  \citenamefont {Dabo}, \citenamefont {Corso}, \citenamefont {{de Gironcoli}},
  \citenamefont {Fabris}, \citenamefont {Fratesi}, \citenamefont {Gebauer},
  \citenamefont {Gerstmann}, \citenamefont {Gougoussis}, \citenamefont
  {Kokalj}, \citenamefont {Lazzeri}, \citenamefont {Martin-Samos},
  \citenamefont {Marzari}, \citenamefont {Mauri}, \citenamefont {Mazzarello},
  \citenamefont {Paolini}, \citenamefont {Pasquarello}, \citenamefont
  {Paulatto}, \citenamefont {Sbraccia}, \citenamefont {Scandolo}, \citenamefont
  {Sclauzero}, \citenamefont {Seitsonen}, \citenamefont {Smogunov},
  \citenamefont {Umari},\ and\ \citenamefont
  {Wentzcovitch}}]{Giannozzi09p395502a}%
  \BibitemOpen
  \bibfield  {author} {\bibinfo {author} {\bibfnamefont {P.}~\bibnamefont
  {Giannozzi}}, \bibinfo {author} {\bibfnamefont {S.}~\bibnamefont {Baroni1}},
  \bibinfo {author} {\bibfnamefont {N.}~\bibnamefont {Bonini}}, \bibinfo
  {author} {\bibfnamefont {M.}~\bibnamefont {Calandra}}, \bibinfo {author}
  {\bibfnamefont {R.}~\bibnamefont {Car}}, \bibinfo {author} {\bibfnamefont
  {C.}~\bibnamefont {Cavazzoni}}, \bibinfo {author} {\bibfnamefont
  {D.}~\bibnamefont {Ceresoli}}, \bibinfo {author} {\bibfnamefont {G.~L.}\
  \bibnamefont {Chiarotti}}, \bibinfo {author} {\bibfnamefont {M.}~\bibnamefont
  {Cococcioni}}, \bibinfo {author} {\bibfnamefont {I.}~\bibnamefont {Dabo}},
  \bibinfo {author} {\bibfnamefont {A.~D.}\ \bibnamefont {Corso}}, \bibinfo
  {author} {\bibfnamefont {S.}~\bibnamefont {{de Gironcoli}}}, \bibinfo
  {author} {\bibfnamefont {S.}~\bibnamefont {Fabris}}, \bibinfo {author}
  {\bibfnamefont {G.}~\bibnamefont {Fratesi}}, \bibinfo {author} {\bibfnamefont
  {R.}~\bibnamefont {Gebauer}}, \bibinfo {author} {\bibfnamefont
  {U.}~\bibnamefont {Gerstmann}}, \bibinfo {author} {\bibfnamefont
  {C.}~\bibnamefont {Gougoussis}}, \bibinfo {author} {\bibfnamefont
  {A.}~\bibnamefont {Kokalj}}, \bibinfo {author} {\bibfnamefont
  {M.}~\bibnamefont {Lazzeri}}, \bibinfo {author} {\bibfnamefont
  {L.}~\bibnamefont {Martin-Samos}}, \bibinfo {author} {\bibfnamefont
  {N.}~\bibnamefont {Marzari}}, \bibinfo {author} {\bibfnamefont
  {F.}~\bibnamefont {Mauri}}, \bibinfo {author} {\bibfnamefont
  {R.}~\bibnamefont {Mazzarello}}, \bibinfo {author} {\bibfnamefont
  {S.}~\bibnamefont {Paolini}}, \bibinfo {author} {\bibfnamefont
  {A.}~\bibnamefont {Pasquarello}}, \bibinfo {author} {\bibfnamefont
  {L.}~\bibnamefont {Paulatto}}, \bibinfo {author} {\bibfnamefont
  {C.}~\bibnamefont {Sbraccia}}, \bibinfo {author} {\bibfnamefont
  {S.}~\bibnamefont {Scandolo}}, \bibinfo {author} {\bibfnamefont
  {G.}~\bibnamefont {Sclauzero}}, \bibinfo {author} {\bibfnamefont {A.~P.}\
  \bibnamefont {Seitsonen}}, \bibinfo {author} {\bibfnamefont {A.}~\bibnamefont
  {Smogunov}}, \bibinfo {author} {\bibfnamefont {P.}~\bibnamefont {Umari}}, \
  and\ \bibinfo {author} {\bibfnamefont {R.~M.}\ \bibnamefont {Wentzcovitch}},\
  }\href@noop {} {\bibfield  {journal} {\bibinfo  {journal} {J. Phys.: Condens.
  Matter}\ }\textbf {\bibinfo {volume} {21}},\ \bibinfo {pages} {395502}
  (\bibinfo {year} {2009})}\BibitemShut {NoStop}%
\bibitem [{\citenamefont {Anisimov}\ \emph {et~al.}(1991)\citenamefont
  {Anisimov}, \citenamefont {Zaanen},\ and\ \citenamefont
  {Andersen}}]{Anisimov91p943}%
  \BibitemOpen
  \bibfield  {author} {\bibinfo {author} {\bibfnamefont {V.~I.}\ \bibnamefont
  {Anisimov}}, \bibinfo {author} {\bibfnamefont {J.}~\bibnamefont {Zaanen}}, \
  and\ \bibinfo {author} {\bibfnamefont {O.~K.}\ \bibnamefont {Andersen}},\
  }\href@noop {} {\bibfield  {journal} {\bibinfo  {journal} {Phys. Rev. B}\
  }\textbf {\bibinfo {volume} {44}},\ \bibinfo {pages} {943} (\bibinfo {year}
  {1991})}\BibitemShut {NoStop}%
\bibitem [{\citenamefont {Hamann}\ \emph {et~al.}(2006)\citenamefont {Hamann},
  \citenamefont {Muller},\ and\ \citenamefont {Hwang}}]{Hamann06p195403}%
  \BibitemOpen
  \bibfield  {author} {\bibinfo {author} {\bibfnamefont {D.~R.}\ \bibnamefont
  {Hamann}}, \bibinfo {author} {\bibfnamefont {D.~A.}\ \bibnamefont {Muller}},
  \ and\ \bibinfo {author} {\bibfnamefont {H.~Y.}\ \bibnamefont {Hwang}},\
  }\href {\doibase 10.1103/PhysRevB.73.195403} {\bibfield  {journal} {\bibinfo
  {journal} {Phys. Rev. B}\ }\textbf {\bibinfo {volume} {73}},\ \bibinfo
  {pages} {195403} (\bibinfo {year} {2006})}\BibitemShut {NoStop}%
\bibitem [{\citenamefont {Baldereschi}\ \emph {et~al.}(1988)\citenamefont
  {Baldereschi}, \citenamefont {Baroni},\ and\ \citenamefont
  {Resta}}]{Baldereschi88p734}%
  \BibitemOpen
  \bibfield  {author} {\bibinfo {author} {\bibfnamefont {A.}~\bibnamefont
  {Baldereschi}}, \bibinfo {author} {\bibfnamefont {S.}~\bibnamefont {Baroni}},
  \ and\ \bibinfo {author} {\bibfnamefont {R.}~\bibnamefont {Resta}},\
  }\href@noop {} {\bibfield  {journal} {\bibinfo  {journal} {Phys. Rev. Lett.}\
  }\textbf {\bibinfo {volume} {61}},\ \bibinfo {pages} {734} (\bibinfo {year}
  {1988})}\BibitemShut {NoStop}%
\bibitem [{\citenamefont {Chen}\ \emph
  {et~al.}(2010{\natexlab{a}})\citenamefont {Chen}, \citenamefont {Kolpak},\
  and\ \citenamefont {Ismail-Beigi}}]{Chen10p85430}%
  \BibitemOpen
  \bibfield  {author} {\bibinfo {author} {\bibfnamefont {H.}~\bibnamefont
  {Chen}}, \bibinfo {author} {\bibfnamefont {A.}~\bibnamefont {Kolpak}}, \ and\
  \bibinfo {author} {\bibfnamefont {S.}~\bibnamefont {Ismail-Beigi}},\ }\href
  {\doibase 10.1103/PhysRevB.82.085430} {\bibfield  {journal} {\bibinfo
  {journal} {Phys. Rev. B}\ }\textbf {\bibinfo {volume} {82}},\ \bibinfo
  {pages} {085430} (\bibinfo {year} {2010}{\natexlab{a}})}\BibitemShut
  {NoStop}%
\bibitem [{\citenamefont {Chen}\ \emph
  {et~al.}(2010{\natexlab{b}})\citenamefont {Chen}, \citenamefont {Kolpak},\
  and\ \citenamefont {Ismail-Beigi}}]{Chen10p2881}%
  \BibitemOpen
  \bibfield  {author} {\bibinfo {author} {\bibfnamefont {H.}~\bibnamefont
  {Chen}}, \bibinfo {author} {\bibfnamefont {A.~M.}\ \bibnamefont {Kolpak}}, \
  and\ \bibinfo {author} {\bibfnamefont {S.}~\bibnamefont {Ismail-Beigi}},\
  }\href {\doibase 10.1002/adma.200903800} {\bibfield  {journal} {\bibinfo
  {journal} {Adv. Mater.}\ }\textbf {\bibinfo {volume} {22}},\ \bibinfo {pages}
  {2881} (\bibinfo {year} {2010}{\natexlab{b}})}\BibitemShut {NoStop}%
\bibitem [{\citenamefont {Wang}\ \emph {et~al.}(2010)\citenamefont {Wang},
  \citenamefont {Niranjan}, \citenamefont {Janicka}, \citenamefont {Velev},
  \citenamefont {Zhuravlev}, \citenamefont {Jaswal},\ and\ \citenamefont
  {Tsymbal}}]{Wang10p094114}%
  \BibitemOpen
  \bibfield  {author} {\bibinfo {author} {\bibfnamefont {Y.}~\bibnamefont
  {Wang}}, \bibinfo {author} {\bibfnamefont {M.~K.}\ \bibnamefont {Niranjan}},
  \bibinfo {author} {\bibfnamefont {K.}~\bibnamefont {Janicka}}, \bibinfo
  {author} {\bibfnamefont {J.~P.}\ \bibnamefont {Velev}}, \bibinfo {author}
  {\bibfnamefont {M.~Y.}\ \bibnamefont {Zhuravlev}}, \bibinfo {author}
  {\bibfnamefont {S.~S.}\ \bibnamefont {Jaswal}}, \ and\ \bibinfo {author}
  {\bibfnamefont {E.~Y.}\ \bibnamefont {Tsymbal}},\ }\href@noop {} {\bibfield
  {journal} {\bibinfo  {journal} {Phys. Rev. B}\ }\textbf {\bibinfo {volume}
  {82}},\ \bibinfo {pages} {094114} (\bibinfo {year} {2010})}\BibitemShut
  {NoStop}%
\end{thebibliography}%

\end{document}